# Citation advantage of COVID-19 related publications

Weishu Liu    wsliu08@163.com
https://orcid.org/0000-0001-8780-6709
School of Information Management, Zhejiang University of Finance and Economics, Hangzhou 310018, Zhejiang, China

Xuping Huangfu    huangfuxupingzufe@163.com
School of Information Management, Zhejiang University of Finance and Economics, Hangzhou 310018, Zhejiang, China

Haifeng Wang    corresponding author    wanghf@shisu.edu.cn
https://orcid.org/0000-0003-4967-7971
School of Business and Management, Shanghai International Studies University, Shanghai, China

**Abstract:** With the global spread of the COVID-19 pandemic, scientists from various disciplines responded quickly to this historical public health emergency. The sudden boom of COVID-19 related papers in a short period of time may bring unexpected influence to some commonly used bibliometric indicators. By a large-scale investigation using Science Citation Index Expanded and Social Sciences Citation Index, this brief communication confirms the citation advantage of COVID-19 related papers empirically through the lens of Essential Science Indicators' highly cited paper. More than 8% of COVID-19 related papers published during 2020 and 2021 were selected as Essential Science Indicators highly cited papers, which was much higher than the set global benchmark value of 1%. The citation advantage of COVID-19 related papers for different Web of Science categories/countries/journal impact factor quartiles were also demonstrated. The distortions of COVID-19 related papers' citation advantage to some bibliometric indicators such as journal impact factor were discussed at the end of this brief communication.

## Introduction

With the global spread of the COVID-19 pandemic, worldwide researchers from various disciplines responded rapidly to this historical public health emergency [1]. One piece of evidence is the sudden boom of COVID-19 related research publications reported by many studies via using different data sources [2-6]. Facing this disruptive event, Nature also investigated "how a torrent of COVID science changed research publishing — in seven charts" [7]. More specifically, Cai et al. [8] documented this historic spectacle as follows:

> "We find that the number of coronavirus publications has seen a great boom in 2020, rising at a spectacular rate from a total of 4,875 articles produced on the topic (preprint and peer reviewed) between January and mid-April to an overall sum of 44,013 by mid-July, and 87,515 by the start of October 2020 (in comparison, nanoscale science was a rapidly growing field in the 1990s, but it took more than 19 years to go from 4000 to 90,000 articles (Grieneisen and Zhang, 2011))" [9].

To quickly respond to the needs of the COVID-19 pandemic, academic publishers have worked with the scientific community to shorten the time to the acceptance of COVID-19 related manuscripts [10]. Preprint platforms such as medRxiv and bioRxiv also play an important role in the timely dissemination of COVID-19 related scientific research results [11-13]. However, there are some downsides to the quick response from the scientific community such as misinformation about COVID-19 and corrections/retractions to published COVID-19 related papers [14-16]. The COVID-19 pandemic may also result in a squeeze on research funding and a possible crowding-out effect on research in other areas [10, 17].

The sudden boom of COVID-19 related papers in a short period of time may also bring unexpected influence to some commonly used bibliometric indicators [2, 18]. For example, by using Scopus, Ioannidis et al. [19] found "massive covidization of research citations and the citation elite" in recent two years. By using 24 major scientific journals, Brandt et al. [20] recently found that COVID-19 papers received 80% more citations than non-COVID-19 papers. A latest short research note even reported that COVID-19 related papers contributed 50% on the journal impact factor of high impact medicine journals [21]. Previous studies suggested that COVID-19 related papers may pose serious challenges to traditional bibliometric indicators. Research field and publication year normalized highly cited paper (HCP) is also an important and widely-used bibliometric indicator. This normalized indicator is more suitable to draw a more comprehensive picture of the citation advantage of all COVID-19 related papers by taking all disciplines into consideration. In this brief communication, we try to probe the citation patterns of COVID-19 related papers via this angle: are COVID-19 related papers more likely to be highly cited papers? We will also discuss the distortions of COVID-19 related papers' citation advantage to some bibliometric indicators at the end of this brief communication.

## Data and methods

Scopus and Web of Science Core Collection are two authoritative and widely used bibliometric databases [22]. In this study, we chose Science Citation Index Expanded (SCIE) and Social Sciences Citation Index (SSCI) of Web of Science Core Collection to probe the citation advantage of COVID-19 related papers. Highly cited papers (HCPs) provided in Incites Essential Science Indicators (ESI) were adopted in this investigation (InCites Essential Science Indicators are updated every two months. The ESI data used in this brief study were updated on July 14, 2022. For more information, please refer to http://esi.help.clarivate.com/Content/scope-coverage.htm). The HCP provided by ESI is a research field (22 broad research fields) and publication year normalized indicator.

With reference to previous studies, the following search query was used to retrieve COVID-19 related papers [23, 24] (Some similar search queries for COVID-19 related publications also exist in existing literature. However, different search queries may return similar numbers of papers since many COVID-19 related papers share one or more commonly used COVID-19 related keywords. For example, by taking some less commonly used or non-standard keywords into the search query, Hook et al.'s search query can return only two new papers which cannot be hit by the search query of this study [25]. Although

Malekpour et al.'s search query can return 591 new papers, they only account for about 0.49% of the COVID-19 related papers identified by this study [26]. The inclusion of some less commonly used or non-standard keywords into the search query will not change the main results of this article. Besides, to search in the topic field (field tag: TS) of Web of Science Core Collection may also introduce some less relevant literature [27]. These two points are two acceptable deficiencies of this study.). Only articles and review articles were considered in this study. Since Web of Science Core Collection has begun to index early access contents in recent years, we used final publication year field (field tag: FPY) to retrieve records published during 2020-2021 (According to the publication year identification rules of ever-early access contents in Web of Science [28], a limited number of records were assigned to the publication year 2019 or 2022 on the search results page of Web of Science Core Collection platform.). The journal impact factor quartiles (JIF quartiles) for corresponding journals were adopted from the latest 2022 Journal Citation Reports. Data were accessed on August 9, 2022.

> TS=("COVID-19" OR "2019-nCoV" OR Coronavirus OR "Corona virus" OR "SARS-CoV" OR "MERS-CoV" OR "Severe Acute Respiratory Syndrome" OR "Middle East Respiratory Syndrome") AND FPY=2020-2021
> Citation index=SCIE, SSCI
> Document type=Article, Review article

## Over-representation of ESI HCPs among COVID-19 related papers

Incites Essential Science Indicators selected the top 1% highly cited articles and review articles in each of the 22 broad research fields and a given year as HCPs. Using SCIE and SSCI, we identified 4,246,285 articles and review articles published during 2020 and 2021. For these records, about 1.05% of them (44,401 papers) were identified as HCPs (a bit higher than the set benchmark value). Surprisingly, for the 119,399 COVID-19 related papers published during the same period, 8.54% of them (10,198) were HCPs which was much higher than the set global benchmark value of 1%. Although the COVID-19 related papers accounted for about 2.81% of all SCIE/SSCI papers, 22.97% of the HCPs indexed by SCIE/SSCI for the same period were identified as COVID-19 related. For the remaining 97.19% of the SCIE/SSCI papers, only 0.83% of them can be selected as ESI HCPs. In other words, COVID-19 related papers have a crowding-out effect on other SCIE/SSCI papers in becoming ESI HCPs.

## Citation advantage for different Web of Science categories

Since Web of Science categories are different from 22 broad research fields of ESI, the proportions of HCPs among all SCIE/SSCI papers for different Web of Science categories will fluctuate around 1%. The last column of Table 1 showed the proportions of HCPs among all SCIE/SSCI articles and reviews for the main categories involving COVID-19 related papers.

Echoing many previous studies [13, 23], in addition to the medical related categories, researchers in many other fields around the world have also carried out numerous COVID-19 related research. Data showed that COVID-19 related papers covered over 200 Web of Science categories. Table 1 showed the distribution of COVID-19 related papers among the top 20 productive categories. For the main categories with a large number of related papers, their proportions of HCPs among COVID-19 related papers were much higher than the corresponding benchmark values for all their SCIE/SSCI papers. Public, Environmental & Occupational Health was the category with the largest number of COVID-19 related records. For 101,598 of all SCIE/SSCI indexed articles and reviews published in this category, 1323

(1.30%) of them were HCPs. Surprisingly, 7.56% (974 out of 12,889) of COVID-19 related papers published in this category were HCPs (the crowding-out effect was also very evident for this category: only 0.39 % of other papers in this category could be selected as ESI HCPs). For less related categories such as Operations Research & Management Science, 14.73% of COVID-19 related papers in this category were HCPs, which was also much higher than the average HCPs rate of 1.27% for all SCIE/SSCI papers in this category.

## Citation advantage for different countries/regions

Some developed countries such as the United States have citation advantages over some developing countries [29-32]. A large proportion of contributions from some developed countries may be a cause for the over-representation of ESI HCPs among COVID-19 related papers. We listed the top 20 countries/regions which contributed the most to COVID-19 related papers. The proportions of HCPs among COVID-19 related papers and all SCIE/SSCI articles/reviews of these top countries/regions were also provided in Table 2.

Supporting previous studies [33-35], Table 2 showed that China surpassed the USA regarding the number of SCIE/SSCI indexed articles and reviews published during 2020 and 2021. However, USA was still the largest contributor of COVID-19 related papers with 35,437 papers. China was the second largest contributor with 17,167 COVID-19 related papers (According to Cai et al. [8] and Wagner et al. [36], China took an early lead on COVID-19 research, but was overtaken by the United States during the second half of 2020.). The UK and Italy were another two large contributors each with over 10,000 related papers.

For the productive countries/regions, their proportions of HCPs among all their articles and reviews in SCIE/SSCI differed. For example, only 0.87% of articles and reviews in SCIE/SSCI contributed by Brazil were HCPs, which was lower than the global benchmark value of 1%. However, the corresponding proportion for the UK was 2.12%. Similarly, these productive countries/regions' proportions of HCPs among all their COVID-19 related papers also varied. For example, 8.09 % of COVID-19 related papers contributed by Brazil were HCPs and the corresponding proportion for the UK was 12.86%. Evidently, for productive countries/regions, their proportions of HCPs among COVID-19 related papers were much higher than corresponding proportions among all articles and reviews in SCIE/SSCI.

Table 1 Proportions of ESI HCPs among top categories

| Dataset | | COVID-19 related articles and reviews | | | All articles and reviews in SCIE/SSCI | | |
|---|---|---|---|---|---|---|---|
| Rank | Web of Science categories | Record count | HCPs | Proportion of HCP (%) | Record count | HCPs | Proportion of HCPs (%) |
| 1 | Public Environmental Occupational Health | 12,889 | 974 | 7.56 | 101,598 | 1,323 | 1.30 |
| 2 | Medicine General Internal | 10,316 | 930 | 9.02 | 92,438 | 1,666 | 1.80 |
| 3 | Environmental Sciences | 7,389 | 544 | 7.36 | 221,261 | 3,258 | 1.47 |
| 4 | Infectious Diseases | 7,265 | 554 | 7.63 | 40,975 | 621 | 1.52 |
| 5 | Multidisciplinary Sciences | 7,006 | 810 | 11.56 | 151,973 | 3,227 | 2.12 |
| 6 | Immunology | 6,726 | 468 | 6.96 | 64,101 | 609 | 0.95 |
| 7 | Pharmacology Pharmacy | 4,840 | 432 | 8.93 | 113,797 | 1,236 | 1.09 |
| 8 | Medicine Research Experimental | 4,583 | 456 | 9.95 | 75,285 | 827 | 1.10 |
| 9 | Microbiology | 4,302 | 446 | 10.37 | 63,231 | 611 | 0.97 |
| 10 | Biochemistry Molecular Biology | 3,903 | 496 | 12.71 | 156,972 | 1,752 | 1.12 |
| 11 | Health Care Sciences Services | 3,821 | 201 | 5.26 | 38,271 | 303 | 0.79 |
| 12 | Psychiatry | 3,726 | 475 | 12.75 | 51,535 | 825 | 1.60 |
| 13 | Virology | 3,317 | 264 | 7.96 | 15,504 | 279 | 1.80 |
| 14 | Surgery | 2,704 | 157 | 5.81 | 84,801 | 303 | 0.36 |
| 15 | Clinical Neurology | 2,688 | 332 | 12.35 | 71,017 | 687 | 0.97 |
| 16 | Pediatrics | 2,590 | 185 | 7.14 | 42,431 | 224 | 0.53 |
| 17 | Chemistry Multidisciplinary | 2,486 | 133 | 5.35 | 205,860 | 3,624 | 1.76 |
| 18 | Cardiac Cardiovascular Systems | 2,306 | 240 | 10.41 | 48,929 | 674 | 1.38 |
| 19 | Cell Biology | 2,220 | 422 | 19.01 | 74,391 | 1,063 | 1.43 |
| 20 | Environmental Studies | 2,129 | 86 | 4.04 | 48,951 | 578 | 1.18 |

Table 2 Proportions of ESI HCPs among productive countries/regions

| Dataset | | COVID-19 related articles and reviews | | | All articles and reviews in SCIE/SSCI | | |
|---|---|---|---|---|---|---|---|
| Rank | Countries/Regions | Record count | HCPs | Proportion of HCPs (%) | Record count | HCPs | Proportion of HCPs (%) |
| 1 | USA | 35,437 | 4,072 | 11.49 | 955,100 | 15,194 | 1.59 |
| 2 | Peoples R China | 17,167 | 1,888 | 11.00 | 1,148,013 | 17,146 | 1.49 |
| 3 | UK | 12,978 | 1,669 | 12.86 | 308,681 | 6,544 | 2.12 |
| 4 | Italy | 10,752 | 1,145 | 10.65 | 195,257 | 3,602 | 1.84 |
| 5 | Germany | 6,765 | 758 | 11.20 | 270,827 | 4,357 | 1.61 |
| 6 | India | 6,226 | 464 | 7.45 | 200,383 | 2,334 | 1.16 |
| 7 | Spain | 6,047 | 599 | 9.91 | 161,187 | 2,610 | 1.62 |
| 8 | Canada | 5,936 | 741 | 12.48 | 175,701 | 3,259 | 1.85 |
| 9 | Australia | 5,293 | 599 | 11.32 | 172,002 | 3,769 | 2.19 |
| 10 | France | 4,743 | 611 | 12.88 | 174,370 | 2,966 | 1.70 |
| 11 | Brazil | 4,029 | 326 | 8.09 | 126,748 | 1,108 | 0.87 |
| 12 | Turkey | 3,348 | 188 | 5.62 | 83,612 | 1,084 | 1.30 |
| 13 | Saudi Arabia | 2,965 | 186 | 6.27 | 62,602 | 1,412 | 2.26 |
| 14 | Japan | 2,953 | 219 | 7.42 | 191,915 | 1,919 | 1.00 |
| 15 | Switzerland | 2,808 | 383 | 13.64 | 80,095 | 1,922 | 2.40 |
| 16 | Netherlands | 2,800 | 427 | 15.25 | 102,843 | 2,297 | 2.23 |
| 17 | South Korea | 2,784 | 194 | 6.97 | 153,066 | 1,800 | 1.18 |
| 18 | Iran | 2,725 | 234 | 8.59 | 98,100 | 1,423 | 1.45 |
| 19 | Poland | 2,088 | 136 | 6.51 | 81,489 | 812 | 1.00 |
| 20 | Belgium | 1,866 | 256 | 13.72 | 57,283 | 1,317 | 2.30 |

## Citation advantage for different JIF quartiles

Generally, papers published in high impact factor journals can attract more citations [37]. We further probed the citation advantage of COVID-19 related papers through the lens of JIF quartile. The latest 2022 Journal Citation Reports were used to retrieve the JIF quartile values for journals which published COVID-19 related papers. If one journal belongs to multiple Web of Science categories, it may have different JIF quartile values [38, 39]. In this study, we chose the highest JIF quartile as this journal's JIF quartile (Q1 is the highest JIF quartile and Q4 is the lowest JIF quartile).

Table 3 listed the distribution of COVID-19 related papers among four JIF quartiles. A limited number of papers which have no matched JIF quartile values were excluded from the analysis in this section. According to Liu et al. [40] and Liu et al. [41], generally, much more than 25% of SCIE/SSCI papers were published in JIF Q1 journals and much less than 25% of them were published in JIF Q4 journals. About 43.80%/32.45%/14.15%/9.60% of COVID-19 related papers were published in JIF Q1/Q2/Q3/Q4 journals respectively. In contrast, for the COVID-19 related ESI HCPs, about 77.87%/17.44%/3.90%/0.79% of them were published in JIF Q1/Q2/Q3/Q4 journals respectively. According to Table 3, 15.24% of COVID-19 related papers which were published in JIF Q1 journals were identified as ESI HCPs. Comparatively, only 4.61%/2.36%/0.71% of the COVID-19 related papers published in JIF Q2/Q3/Q4 journals were ESI HCPs respectively. For example, about 77.44% of COVID-

19 related papers published in *Science* (JIF Q1 journal) have been selected as ESI HCPs; comparatively, only about 0.60% of COVID-19 related papers published in *Journal of Infection in Developing Countries* (JIF Q4 journal) have been selected as ESI HCPs. We can conclude that COVID-19 related ESI HCPs mainly concentrated in journals of high JIF quartiles (especially Q1) and COVID-19 related papers which were published in high JIF quartile journals had a much higher probability of becoming ESI HCPs.

Table 3 Proportions of ESI HCPs among four JIF quartiles

| JIF quartiles | COVID-19 related papers | Share (%) | COVID-19 related HCPs | Share (%) | Proportion of HCPs (%) |
|---|---|---|---|---|---|
| Q1 | 52,070 | 43.80 | 7,936 | 77.87 | 15.24 |
| Q2 | 38,579 | 32.45 | 1,777 | 17.44 | 4.61 |
| Q3 | 16,817 | 14.15 | 397 | 3.90 | 2.36 |
| Q4 | 11,408 | 9.60 | 81 | 0.79 | 0.71 |
| Total | 118,874 | 100.00 | 10,191 | 100.00 | 8.57 |

## Discussion

Worldwide researchers from various disciplines responded quickly to this historical public health emergency caused by COVID-19, which was further evidenced by the staggering increase in the number of COVID-19 related research papers [7, 8, 24, 42]. Using SCIE/SSCI indexed papers published during 2020 and 2021, we probed the COVID-19 papers' citation advantage empirically through the lens of ESI HCPs. Data showed that 8.54% of COVID-19 related papers were ESI HCPs which was much higher than the set global benchmark value of 1%. Meanwhile, the boom of COVID-19 related papers in a short period of time also introduced the crowding-out effect: only 0.83% of other SCIE/SSCI papers could be selected as ESI HCPs. The crowding-out effect was very evident for some categories such as Public, Environmental & Occupational Health.

COVID-19 related papers covered over 200 Web of Science categories. For the main categories with a large number of related papers, their proportions of HCPs among their COVID-19 related papers were much higher than the corresponding benchmark values. For productive countries/regions, various proportions of their COVID-19 related papers were ESI HCPs. However, their proportions of ESI HCPs among COVID-19 related papers were also much higher than corresponding values for all articles and reviews in SCIE/SSCI. COVID-19 related HCPs mainly concentrated in journals of high JIF quartiles and COVID-19 related papers which published in high JIF quartile journals had a much higher probability of becoming ESI HCPs.

The citation advantage of COVID-19 related papers may be temporary and can be explained via two different angles. On the one hand, the huge demand for references to the existing newly published COVID-19 related literature exploded in a short period of time. The booming number of COVID-19 related papers themselves also need to cite a large number of recently published studies. Many less relevant studies conducted in the context of COVID-19 (this kind of studies may not be identified as COVID-19 related papers by the method used in this study) also need to cite COVID-19 related papers as the background material. On the other hand, the citation thresholds for newly published papers (for example published in 2020 or 2021) to be selected as ESI HCPs are relatively low. One paper can easily become an ESI HCP if it is cited ten or twenty times within one or two years, which is relatively common under the COVID-19 pandemic.

Similar to the useful but controversial indicator Journal impact factor, ESI HCP is also a useful indicator in research evaluation. Scholars can also use this indicator to identify influential research. The citation advantage of COVID-19 related papers also means that it's harder for other papers to be selected as ESI HCPs (namely crowding-out effect). Therefore, some functions of ESI HCPs such as influential research identification and ESI discipline ranking will be disrupted. The citation advantage of COVID-19 related papers will also benefit the corresponding journals' JIF values especially for some small journals [2, 21, 43, 44]. One typical example is the leap of the journal impact factor of Journal of Public Transportation in the latest Journal Citation Reports. The JIF of this journal rose from 2.529 (ranked as 27/37 in the category of Transportation-SSCI) in 2020 to 37.667 in 2021(ranked as 1/37 in the category of Transportation-SSCI). Through further exploration of the 2021 JIF calculation, we found that this journal had only three citable items and the latest JIF of this journal was almost completely contributed by one COVID-19 related paper with 112 citations. Fortunately, as the popularity of COVID-19 research gradually fades, the disturbance of COVID-19 related papers on various bibliometric indicators such as journal impact factor will also gradually weaken. In practice, however, we re-emphasize the need for judicious use of bibliometric indicators, as emphasized by Clarivate and the Leiden Manifesto [45].

This brief empirical investigation also has some limitations. Firstly, the retrieval strategy used in this brief communication may over-estimate the number of COVID-19 related publications (Even if the used search strategy also omits a small share of papers that only use uncommon or non-standard COVID-19 keywords) and further under-estimate the citation advantage of COVID-19 related publications. Secondly, this study only uses two citation indexes of Web of Science Core Collection as the data sources, regional databases and preprint platforms may also be considered as additional data sources to draw a more comprehensive picture of citation advantage [12, 36]. Thirdly, we only use the field normalized indicator ESI HCP as a proxy of "highly cited". The characteristics of COVID-19 related papers with high absolute numbers of citations and the impact of these highly cited papers on various bibliometric metrics are also worthy of further investigation. Lastly, the citation advantage of COVID-19 related papers can be more precisely measured by a rigorous econometric model, which includes controlling for factors such as international cooperation and journal quality. In addition to the above four deficiencies, this brief empirical study can provide readers with some interesting features of COVID-19 related papers and shed light on its possible impact on research evaluation and academic publishing.

**Compliance with ethical standards**

**Conflict of interest** The authors declare that they have no conflict of interest.

**Supplemental material**

The full list of Table 1 for this article is available online.